\documentclass[12pt,english]{article}
\usepackage[T1]{fontenc}
\usepackage[latin9]{inputenc}
\usepackage[a4paper]{geometry}
\usepackage{textcomp}
\usepackage{amsmath}
\usepackage{setspace}
\usepackage{amssymb}
\makeatletter

\newcommand{\lyxmathsym}[1]{\ifmmode\begingroup\def\b@ld{bold}
  \text{\ifx\math@version\b@ld\bfseries\fi#1}\endgroup\else#1\fi}

\newcommand{\lyxaddress}[1]{
\par {\raggedright #1
\vspace{1.4em}
\noindent\par}
}

\makeatother

\usepackage{babel}

\begin{document}

\title{SPONTANEOUS SYMMETRY BREAKING IN PRESENCE OF ELECTRIC AND MAGNETIC
CHARGES}

\author{Pushpa, P. S. Bisht and O. P. S. Negi%
\thanks{Present Address from November 08- December22, 2010:\textbf{ Universität
Konstanz, Fachbereich Physik, Postfach-M 677, D-78457 Konstanz, Germany}%
}}

\maketitle

\lyxaddress{\begin{center}
Department of Physics,\\
 Kumaun University, S. S. J. Campus, \\
Almora-263601 (Uttarakhand) India
\par\end{center}}

\lyxaddress{\begin{center}
Email-pushpakalauni60@yahoo.co.in\\
 ps\_bisht 123@rediffmail.com \\
ops\_negi@yahoo.co.in
\par\end{center}}
\begin{abstract}
Starting with the definition of quaternion gauge theory, we have undertaken
the study of $SU(2)_{e}$$\times SU(2)_{m}\times U(1)_{e}\times U(1)_{m}$
in terms of the simultaneous existence of electric and magnetic charges
along with their Yang - Mills counterparts. As such, we have developed
the gauge theory in terms of four coupling constants associated with
four - gauge symmetry $SU(2)_{e}$$\times SU(2)_{m}\times U(1)_{e}\times U(1)_{m}$.
Accordingly, we have made an attempt to obtain the abelian and non
- Abelian gauge structures for the particles carrying simultaneously
the electric and magnetic charges (namely dyons). Starting from the
Lagrangian density of two $SU(2)\times U(1)$ gauge theories responsible
for the existence of electric and magnetic charges, we have discussed
the consistent theory of spontaneous symmetry breaking and Higgs mechanism
in order to generate the masses. From the symmetry breaking, we have
generated the two electromagnetic fields, the two massive vector $W^{\pm}$
and $Z^{0}$ bosons fields and the Higgs scalar fields.
\end{abstract}

\section{Introduction}

		Symmetry plays the central role in determining its dynamical structure.
The Lagrangian exhibits invariance under $SU(2)\times U(1)$ gauge
transformations for the electroweak interactions. Since the imposition
of local symmetry implies the existence of massless vector particles\cite{key-1},
Higgs mechanism is used for the spontaneous breaking of gauge symmetry
to generate masses for the weak gauge bosons charged as well as neutral
particle\cite{key-2}. If these features of the gauge theory are avoided,
we obtain massive vector bosons and hence the gauge symmetry must
be broken. In the Higgs mechanism a larger symmetry is spontaneously
broken into a smaller symmetry through the vacuum expectation value
of the Higgs field and accordingly gauge bosons become massive. The
simplest way of introducing spontaneous symmetry breakdown is to include
scalar Higgs fields by hand into the Lagrangian\cite{key-3}. Recently,
we \cite{key-4} have made an attempt to develop the quaternionic
formulation of Yang \textendash{} Mill\textquoteright{}s field equations
and octonion reformulation of quantum chromo dynamics (QCD) by taking
magnetic monopole into account \cite{key-5}. It has been shown that
the three quaternion units explain the structure of Yang- Mill\textquoteright{}s
field while the seven octonion units provide the consistent structure
of $SU(3)_{C}$ gauge symmetry of quantum chromo dynamics. Our theory
differs from the quaternion gauge theory of spontaneously symmetry
breaking mechanism already developed by others \cite{key-6,key-7,key-8,key-9}
in terms of gauge groups and methedology adoped by them in different
manners. In this paper, we have extended our previous results to develop
a meaningful gauge theory which may purport a model for massive gauge
particles. Starting with the definition of quaternion gauge theory,
we have undertaken the study of $SU(2)_{e}$$\times SU(2)_{m}\times U(1)_{e}\times U(1)_{m}$
in terms of the simultaneous existence of electric and magnetic charges
along with their Yang - Mills counterparts. As such, we have developed
the gauge theory in terms of four coupling constants associated with
four - gauge symmetry $SU(2)_{e}$$\times SU(2)_{m}\times U(1)_{e}\times U(1)_{m}$.
Accordingly, we have made an attempt to obtain the abelian and non
- Abelian gauge structures for the particles carrying simultaneously
the electric and magnetic charges (namely dyons). Starting from the
Lagrangian density of two $SU(2)\times U(1)$ gauge theories responsible
for the existence of electric and magnetic charges, we have discussed
the consistent theory of spontaneous symmetry breaking and Higgs mechanism
in order to generate the masses. From the symmetry breaking, we have
generated the two electromagnetic fields, the two massive vector $W^{\pm}$
and $Z^{0}$ bosons fields and the Higgs scalar fields. Here, we have
briefly discussed the Higgs mechanism for the case of general non-abelian
local symmetries.

\section{Quaternion Gauge Formalism}

Let $\phi(x)$ be a quaternionic field ($Q$ field) and expressed
\cite{key-4,key-6,key-7,key-8,key-9} as

\begin{align}
\phi= & e_{0}\phi_{0}+e_{j}\phi_{j}\,\,(\forall\, j=1,2,3)\label{eq:1}\end{align}
where $\phi_{0}$ and $\phi_{j}$ are local Hermitian fields and $e_{0}$
and $e_{j}$ are the imaginary basis of \textbf{$Q$ }which satisfy
the following property

\begin{eqnarray}
e_{0}^{2} & =e_{0};\,\, & e_{0}e_{j}=-\delta_{jk}e_{0}+\epsilon_{jkl}e_{l}\,\,\,(\forall j,k,l=1,2,3).\label{eq:2}\end{eqnarray}
In global gauge symmetry , the unitary transformations are independent
of space and time. Accordingly, under $SU(2)$ global gauge symmetry,
the quaternion spinor $\psi$ transforms as 

\begin{eqnarray}
\psi & \longmapsto\psi^{\shortmid}= & U\,\psi\label{eq:3}\end{eqnarray}
where $U$ is $2\times2$ unitary matrix and satisfies 

\begin{eqnarray}
U^{\dagger}U & =UU^{\dagger}=UU^{-1}=U^{-1}U= & 1.\label{eq:4}\end{eqnarray}
On the other hand, the quaternion conjugate spinor transforms as \begin{eqnarray}
\overline{\psi}\longmapsto\overline{\psi^{\shortmid}} & = & \overline{\psi}U^{-1}\label{eq:5}\end{eqnarray}
and hence the combination $\psi\overline{\psi}=\overline{\psi}\psi=\psi\overline{\psi^{\shortmid}}=\overline{\psi^{\shortmid}}\psi$
is an invariant quantity. We may thus write any unitary matrix as 

\begin{eqnarray}
U & = & \exp\left(i\,\hat{H}\right)\,\,(i=\sqrt{-1})\label{eq:6}\end{eqnarray}
where $\hat{H}$ is Hermitian $\hat{H}^{\dagger}=\hat{H}$. Thus,
we express the Hermitian $2\times2$ matrix in terms of four real
numbers, $a_{1,}a_{2},\, a_{3}$ and $\theta$ as 

\begin{eqnarray}
\hat{H} & = & \theta\hat{1}+\tau_{j}a_{j}=\theta\hat{1}+ie_{j}a_{j}\label{eq:7}\end{eqnarray}
where $\hat{1}$ is the $2\times2$ unit matrix, $\tau_{j}$ are well
known $2\times2$ Pauli-spin matrices and $e_{1},\, e_{2},\, e_{3}$
are the quaternion units which are connected to Pauli-spin matrices
as \begin{eqnarray}
e_{0}= & 1;\,\,\,\,\, & e_{j}=-i\tau_{j}.\label{eq:8}\end{eqnarray}
 Hence, we write the Hermitian matrix $\hat{H}$ as 

\begin{eqnarray}
\hat{H} & = & \left(\begin{array}{cc}
\theta+a_{3} & a_{1}-ia_{2}\\
a_{1}+ia_{2} & \theta-a_{3}\end{array}\right).\label{eq:9}\end{eqnarray}
Equation (\ref{eq:6}) is now reduced to be 

\begin{eqnarray}
U & = & \exp\left(i\,\theta\right).\exp\left(-e_{j}a_{j}\right).\label{eq:10}\end{eqnarray}
For $SU(2)$ global gauge transformations both $\theta$ and $\overrightarrow{a}$
are independent of space time. Here $\exp\left(i\,\theta\right)$
describes the $U(1)$ gauge transformations while the term $\exp\left(-e_{j}a_{j}\right)$
represents the non-Abelian $SU(2)$ gauge transformations. Thus under
global $SU(2)$ gauge transformations, the Dirac spinor $\psi$ transforms
as 

\begin{eqnarray}
\psi & \longmapsto\psi^{\shortmid}= & U\,\psi=\exp\left(-e_{j}a_{j}\right)\psi.\label{eq:11}\end{eqnarray}
The generators of this group $e_{j}$ obey the commutation relation;

\begin{eqnarray}
\left[e_{j},e_{k}\right] & = & 2f_{jkl}e_{l}\label{eq:12}\end{eqnarray}
which implies $e_{j}e_{k}\neq e_{k}e_{j}$ showing that the elements
of the group are not commutating giving rise to the non abelian gauge
structure whose structure condtant is $f_{jkl}$. So, the partial
derivative of spinor $\psi$ transforms as 

\begin{eqnarray}
\partial_{\mu}\psi(x) & \longmapsto & \partial_{\mu}\psi^{'}(x)\Longrightarrow\exp\left(-e_{j}a_{j}\right)(\partial_{\mu}\psi).\label{eq:13}\end{eqnarray}
For $SU(2)$ local gauge transformation we may replace the unitary
gauge transformation as space - time dependent. So on replacing $U$
by $S$ in equation (\ref{eq:10}), we get

\begin{eqnarray}
\psi & \longmapsto\psi^{\shortmid}= & S\,\psi\label{eq:14}\end{eqnarray}
in which \begin{eqnarray}
S & = & \exp[-\sum_{j}\, q\, e_{j}\zeta_{j}(x)]\label{eq:15}\end{eqnarray}
where parameter $\overrightarrow{\zeta}=-\frac{\overrightarrow{a}(x)}{q}$
with $\overrightarrow{a}(x)$ is infinitesimal quantity depends on
space while time and $q$ is described as the coupling constant. In
$SU(2)$ local gauge symmetry as the partial derivative which contains
an extra term is then replaced by a covariant derivative i.e. 

\begin{eqnarray}
\partial_{\mu}\psi\mapsto S\partial_{\mu}\psi & +\left(\partial_{\mu}S\right)\psi & \longmapsto D_{\mu}\psi\label{eq:16}\end{eqnarray}
where the covariant derivative $D_{\mu}$ is defined in terms of two
$Q$ - gauge fields \cite{key-4} i.e 

\begin{eqnarray}
D_{\mu}\psi & =\partial_{\mu}\psi & +A_{\lyxmathsym{\textmu}}\psi+B_{\lyxmathsym{\textmu}}\psi\label{eq:17}\end{eqnarray}
where $A_{\mu}=-iA_{\mu}^{j}\tau_{j}=A_{\mu}^{j}e_{j}=\overrightarrow{A_{\mu}}\cdot\overrightarrow{e}$
and $B_{\mu}=-iB_{\mu}^{j}\tau_{j}=B_{\mu}^{j}e_{j}=\overrightarrow{B_{\mu}}\cdot\overrightarrow{e}$.
Two gauge fields $A_{\mu}$ and $B_{\mu}$are respectively associated
with electric and magnetic charges of dyons (i.e particles carrying
the simultaneous existence of electric and magnetic charges). Thus
the gauge field $\left\{ A_{\mu}\right\} $ is coupled with the electric
charge while the gauge field $\left\{ B_{\mu}\right\} $ is coupled
with the magnetic charge (i.e. magnetic monopole). These two gauge
fields are subjected by the following gauge transformations

\begin{equation}
A_{\lyxmathsym{\textmu}}'\longmapsto S\, A_{\lyxmathsym{\textmu}}\, S^{-1}+(\partial_{\mu}S)\, S^{-1};\,\,\,\,\: B_{\lyxmathsym{\textmu}}'\longmapsto S\, B_{\lyxmathsym{\textmu}}\, S^{-1}+(\partial_{\mu}S)\, S^{-1}.\label{eq:18}\end{equation}
For the limiting case of infinitesimal transformations of $\zeta$
, we may expand $S$ by keeping only first order terms in the following
manner as

\begin{equation}
S\cong1+\overrightarrow{e}.\overrightarrow{a}(x);\,\,\,\,\,\,\, S^{-1}\cong1-\overrightarrow{e}.\overrightarrow{a}(x);\,\,\,\:\partial_{\mu}\left(S\right)\cong\overrightarrow{e}.\partial_{\mu}\left\{ \overrightarrow{a}\left(x\right)\right\} .\label{eq:19}\end{equation}
Thus, the corresponding gauge fields associated with electric and
magnetic charges are expressed as,

\begin{eqnarray}
A_{\mu}= & g_{e}e_{0}A_{\mu}^{0}+g_{e}^{'}e_{i}A_{\mu}^{i};\,\, & B_{\mu}=g_{m}e_{0}B_{\mu}^{0}+g_{m}^{'}e_{i}B_{\mu}^{i}.\label{eq:20}\end{eqnarray}
In the above equation (\ref{eq:20}) $g_{e}$,\ $g_{m}$,$\, g_{e}^{'}$,$\, g_{m}^{'}$
are the four coupling constants corresponding to the symmetry $U(1)_{e},\, U(1)_{m},\, SU(2)_{e}$,\ $SU(2)_{m}$
so that the gauge fields $A_{\mu}$ and $B_{\mu}$ are transformed
\cite{key-4,key-7} as

\begin{eqnarray}
A_{\mu}^{'}= & UA_{\mu}\overline{U}+U\partial_{\mu}\overline{U}\,\,\, & B_{\mu}^{'}=VB_{\mu}\overline{V}+V\partial_{\mu}\overline{V}.\label{eq:21}\end{eqnarray}
where $U$ and $V$ are two tupes of unitary gauge groups.

\section{Quaternion Spontaneous Symmetry Breaking}

Let us consider the local gauge invariance of the Lagrangian, we get 

\begin{align}
L= & \left(D_{\mu}\phi\right)\left(D^{\mu}\phi\right)^{\star}-V(\phi^{2})-\frac{1}{4}F_{\mu\nu}F^{\mu\nu}-\frac{1}{4}G_{\mu\nu}G^{\mu\nu}-\frac{1}{4}F_{\mu\nu}^{a}F_{a}^{\mu\nu}-\frac{1}{4}G_{\mu\nu}^{a}G_{a}^{\mu\nu}\label{eq:22}\end{align}
where $F_{\mu\nu}$ is the electromagnetic field tensor for the description
of electric charge, ($\star$) is used for complex conjugation when
quaternions are compactified to complex numbers, $G_{\mu\nu}$ is
identical to the dual of $F_{\mu\nu}$and is resposible for the existence
of magnetic charge while the potential term $V(\phi^{2})$ contains
the usual mass and quadretic self - interaction terms described as

\begin{align}
V(\phi^{\dagger}\phi)= & m^{2}\phi^{\dagger}\phi+\lambda\left(\phi^{\dagger}\phi\right)^{2}.\label{eq:23}\end{align}
where $m^{2}$ and $\lambda$ are real constant parameters while the
symbol ($\dagger$) denotes Hermitian conjugation. $\lambda$ should
be positive to ensure the stable vacuum. Furthermore, higher power
terms of $\left(\phi^{\dagger}\phi\right)$ are not allowed in order
to look the theory to be renormalized. Equation (\ref{eq:22}) is
invariant under the local gauge transformations

\begin{align}
\phi^{'}= & \exp\left[ie_{0}\left(g_{e}+g_{m}\right)+e_{j}\left(g_{e}^{'}+g_{m}^{'}\right)\right]\phi.\label{eq:24}\end{align}
Let us take the variation in $\phi$ as,

\begin{align}
\delta\phi= & \left[ie_{0}\left(g_{e}+g_{m}\right)+e_{j}\left(g_{e}^{'}+g_{m}^{'}\right)\right].\label{eq:25}\end{align}
After taking the variations, the Lagrangian (\ref{eq:22}) yields
the following expression for current density $j_{\mu}$ 

\begin{eqnarray}
J_{\mu} & = & \left(j_{\mu}\right)_{U(1)_{e}}+\left(j_{\mu}\right)_{U(1)_{m}}+\left(j_{\mu}\right)_{SU(2)_{e}}+\left(j_{\mu}\right)_{SU(2)_{m}}\label{eq:26}\end{eqnarray}
where

\begin{eqnarray}
\left(j_{\mu}\right)_{U(1)_{e}} & = & ie_{0}g_{e}\left[\phi^{\dagger}D_{\mu}\phi-\phi D_{\mu}\phi^{\dagger}\right];\label{eq:27}\\
\left(j_{\mu}\right)_{U(1)_{m}} & = & ie_{0}g_{m}\left[\phi^{\dagger}D_{\mu}\phi-\phi D_{\mu}\phi^{\dagger}\right];\label{eq:28}\\
\left(j_{\mu}\right)_{SU(2)_{e}} & = & ie_{j}g_{e}^{'}\left[\phi^{\dagger}D_{\mu}\phi-\phi D_{\mu}\phi^{\dagger}\right];\label{eq:29}\\
\left(j_{\mu}\right)_{SU(2)_{m}} & = & ie_{j}g_{m}^{'}\left[\phi^{\dagger}D_{\mu}\phi-\phi D_{\mu}\phi^{\dagger}\right];\label{eq:30}\end{eqnarray}
are the currents respectively associated with the gauge groups $U(1)_{e}$
, $\, U(1)_{m}$, $SU(2)_{e}$ and $SU(2)_{m}$. The condition for
the minimum potential leads to 

\begin{eqnarray}
\frac{\partial V}{\partial\phi} & =0 & \Rightarrow\phi\left(m^{2}+2\lambda\phi^{2}\right).\label{eq:31}\end{eqnarray}
Now may discuss the different cases of $m^{2}$. For $m^{2}>0$ ,
we have the situation for a massive scalar field particle, and $\phi=\phi_{min}$
. Here $\phi=0$, as the vacuum having $V=0$. For the case $m^{2}<0$,
we see that $\phi=\phi_{min}$ and $\phi=\pm v=\pm\left(-\frac{m^{2}}{2\lambda}\right)$
. So, in this case the local gauge symmetry is spontaneously broken
and we may obtain the vacuum expectation value for the scalar field
$\phi$ as,

\begin{eqnarray}
\phi & = & \left[\begin{array}{c}
0\\
\frac{v}{\sqrt{2}}\end{array}\right]\label{eq:32}\end{eqnarray}
with $v=\sqrt{-\frac{m^{2}}{2\lambda}}$ . For particle spectrum,
the vacuum expectation value is modified as 

\begin{eqnarray}
\phi & = & \left[\begin{array}{c}
0\\
\frac{1}{\sqrt{2}}\left(v+\eta(x)\right)\end{array}\right]\label{eq:33}\end{eqnarray}
where $\eta$ is the arbitrary parameter for excited state. Applying
the equation (\ref{eq:33}), Lagrangian (\ref{eq:22}) is modified
as 

\begin{align}
L'= & \frac{1}{2}\partial_{\mu}\eta\partial_{\mu}\eta+\frac{1}{2}.2m^{2}\eta^{2}-\frac{1}{4}F_{\mu\nu}F^{\mu\nu}-\frac{1}{4}G_{\mu\nu}G^{\mu\nu}-\frac{1}{4}F_{\mu\nu}^{a}F_{a}^{\mu\nu}-\frac{1}{4}G_{\mu\nu}^{a}G_{a}^{\mu\nu}-\lambda v\eta^{3}-\frac{\lambda}{4}\eta^{4}\nonumber \\
+ & \frac{1}{2}\left[v+\eta(x)\right]^{2}[g_{e}^{2}\left(A_{\mu}^{0}\right)^{2}+g_{m}^{2}\left(B_{\mu}^{0}\right)^{2}+g_{e}^{'2}\left(A_{\mu}^{j}\right)^{2}+g_{m}^{'2}\left(B_{\mu}^{j}\right)^{2}+2g_{e}g_{m}A_{\mu}^{0}B_{\mu}^{0}+2g_{e}g_{e}^{'}A_{\mu}^{0}A_{\mu}^{j}\nonumber \\
+ & 2g_{e}g_{m}^{'}A_{\mu}^{0}B_{\mu}^{j}+2g_{e}^{'}g_{m}A_{\mu}^{j}B_{\mu}^{0}++2g_{m}g_{m}^{'}B_{\mu}^{0}B_{\mu}^{j}+2g_{e}^{'}g_{m}^{'}A_{\mu}^{j}B_{\mu}^{j}].\label{eq:34}\end{align}
This Lagrangian describes the gauge group $SU(2)_{e}\times SU(2)_{m}\times U(1)_{e}\times U(1)_{m}$.
We see that the Lagrangian density (\ref{eq:34}) is symmetric under
$\eta\rightarrow-\eta$ so that the term $\lambda v\eta^{3}$ breaks
the symmetry. It is to be noted that the Lagrangian density (\ref{eq:22})
is invariant under the changes taken to ground expectation value $\left\langle \phi\right\rangle _{0}=v$
and $\phi=v+\eta,$ but the new Lagrangian (\ref{eq:34}) breakes
the symmetry and is no more invariant under these changes. It is only
due to the mechanism of spontaneous symmetry breaking. Lagrangian
(\ref{eq:34}) is also expressed as

\begin{align}
L'=L_{0}+ & L_{I}\label{eq:35}\end{align}
where $L_{0}$ contains the kinetic energy and mass terms i.e.

\begin{align}
L_{0}= & \frac{1}{2}\partial_{\mu}\eta\partial_{\mu}\eta+\frac{1}{2}.2m^{2}\eta^{2}-\frac{1}{4}F_{\mu\nu}F^{\mu\nu}-\frac{1}{4}G_{\mu\nu}G^{\mu\nu}-\frac{1}{4}F_{\mu\nu}^{a}F_{a}^{\mu\nu}-\frac{1}{4}G_{\mu\nu}^{a}G_{a}^{\mu\nu}+\frac{1}{2}v^{2}[g_{e}^{2}\left(A_{\mu}^{0}\right)^{2}\nonumber \\
+ & g_{m}^{2}\left(B_{\mu}^{0}\right)^{2}+g_{e}^{'2}\left(A_{\mu}^{j}\right)^{2}+g_{m}^{'2}\left(B_{\mu}^{j}\right)^{2}+2g_{e}g_{m}A_{\mu}^{0}B_{\mu}^{0}+2g_{e}g_{e}^{'}A_{\mu}^{0}A_{\mu}^{j}+2g_{e}g_{m}^{'}A_{\mu}^{0}B_{\mu}^{j}\nonumber \\
+ & 2g_{e}g_{e}^{'}B_{\mu}^{0}A_{\mu}^{j}+2g_{m}g_{m}^{'}B_{\mu}^{0}B_{\mu}^{j}+2g_{e}^{'}g_{m}^{'}A_{\mu}^{j}B_{\mu}^{j}]\label{eq:36}\end{align}
and the interaction term becomes,

\begin{align}
L_{I}= & -\lambda v\eta^{3}-\frac{\lambda}{4}\eta^{4}+\left[v\eta+\frac{\eta^{2}}{2}\right][g_{e}^{2}\left(A_{\mu}^{0}\right)^{2}+g_{m}^{2}\left(B_{\mu}^{0}\right)^{2}+g_{e}^{'2}\left(A_{\mu}^{j}\right)^{2}+g_{m}^{'2}\left(B_{\mu}^{j}\right)^{2}+2g_{e}g_{m}A_{\mu}^{0}B_{\mu}^{0}+2g_{e}g_{e}^{'}A_{\mu}^{0}A_{\mu}^{j}\nonumber \\
+ & 2g_{e}g_{m}^{'}A_{\mu}^{0}B_{\mu}^{j}+2g_{e}g_{e}^{'}B_{\mu}^{0}A_{\mu}^{j}+2g_{m}g_{m}^{'}B_{\mu}^{0}B_{\mu}^{j}+2g_{e}^{'}g_{m}^{'}A_{\mu}^{j}B_{\mu}^{j}].\label{eq:37}\end{align}
We may now write the masses for gauge bosons and their scalar interaction
terms in the form of $4\times4$ matrix as,

\begin{align}
\beta= & \left[\begin{array}{cccc}
A_{\mu}^{0} & B_{\mu}^{0} & A_{\mu}^{i} & B_{\mu}^{i}\end{array}\right]\left[\begin{array}{cccc}
g_{e}^{2} & g_{e}g_{m} & g_{e}g_{e}^{'} & g_{e}g_{m}^{'}\\
g_{e}g_{m} & g_{m}^{2} & g_{m}g_{e}^{'} & g_{m}g_{m}^{'}\\
g_{e}g_{e}^{'} & g_{m}g_{e}^{'} & g_{e}^{'2} & g_{e}^{'}g_{m}^{'}\\
g_{e}g_{m}^{'} & g_{m}g_{m}^{'} & g_{e}^{'}g_{m}^{'} & g_{m}^{'2}\end{array}\right]\left[\begin{array}{c}
A_{\mu}^{0}\\
B_{\mu}^{0}\\
A_{\mu}^{i}\\
B_{\mu}^{i}\end{array}\right]\label{eq:38}\end{align}
which can be further reduced to 

\begin{align}
\beta= & g_{e}^{2}\left(A_{\mu}^{0}\right)^{2}+g_{m}^{2}\left(B_{\mu}^{0}\right)^{2}+g_{e}^{'2}\left(A_{\mu}^{j}\right)^{2}+g_{m}^{'2}\left(B_{\mu}^{j}\right)^{2}+2g_{e}g_{m}A_{\mu}^{0}B_{\mu}^{0}+2g_{e}g_{e}^{'}A_{\mu}^{0}A_{\mu}^{j}\nonumber \\
+ & 2g_{e}g_{m}^{'}A_{\mu}^{0}B_{\mu}^{j}+2g_{e}^{'}g_{m}A_{\mu}^{j}B_{\mu}^{0}++2g_{m}g_{m}^{'}B_{\mu}^{0}B_{\mu}^{j}+2g_{e}^{'}g_{m}^{'}A_{\mu}^{j}B_{\mu}^{j}.\label{eq:39}\end{align}

\section{Mass Generation due to Symmetry Breaking}

Now we may discuss the different cases for $SU(2)\times U(1)$ gauge
groups of electro- weak unification as under :

\textbf{Case-I} For usual electroweak unification i.e. $SU(2)_{e}\times U(1)_{e}$,
we assume $g_{m}^{'}=g_{m}=0$ so that the mass contribution term
is given by

\begin{eqnarray}
L_{mass} & = & \left(v\frac{g_{e}^{'}}{2}\right)^{2}W_{\mu}^{+}W_{-}^{\mu}+\frac{v^{2}}{8}\left(\begin{array}{cc}
A_{\mu}^{0} & A_{\mu}^{1}\end{array}\right)\left(\begin{array}{cc}
g_{e}^{2} & g_{e}g_{e}^{'}\\
g_{e}g_{e}^{'} & g_{e}^{'2}\end{array}\right)\left(\begin{array}{c}
A_{\mu}^{0}\\
A_{\mu}^{1}\end{array}\right).\label{eq:40}\end{eqnarray}
It directly gives the masses of charged $W$ bosons as,

\begin{eqnarray}
m_{W} & = & \frac{v}{2}g_{e}^{'};\label{eq:41}\\
W_{\mu}^{\pm} & = & \left(A_{\mu}^{2}\pm e_{3}A_{\mu}^{3}\right);\label{eq:42}\\
Z^{0} & = & \frac{e_{0}g_{e}A_{\mu}^{0}+e_{1}g_{e}^{'}A_{\mu}^{1}}{\sqrt{g_{e}^{2}+g_{e}^{'^{2}}}}.\label{eq:43}\end{eqnarray}

\textbf{Case - II }For magnetoweak unification i.e.$SU(2)_{m}\times U(1)_{m}$,
we assume $g_{e}^{'}=g_{e}=0$ so that the mass contribution term
is given by

\begin{eqnarray}
L_{mass} & = & \left(v\frac{g_{m}^{'}}{2}\right)^{2}W_{\mu}^{+}W_{-}^{\mu}+\frac{v^{2}}{8}\left(\begin{array}{cc}
B_{\mu}^{0} & B_{\mu}^{1}\end{array}\right)\left(\begin{array}{cc}
g_{m}^{2} & g_{m}g_{m}^{'}\\
g_{m}g_{m}^{'} & g_{m}^{'2}\end{array}\right)\left(\begin{array}{c}
B_{\mu}^{0}\\
B_{\mu}^{1}\end{array}\right).\label{eq:44}\end{eqnarray}
Here we have another type of electroweak interaction due to the presence
of magnetic monopole for which the masses for the charged $W$ bosons
are obtained as

\begin{eqnarray}
m_{W} & = & \frac{v}{2}g_{m}^{'};\label{eq:45}\\
W_{\mu}^{\pm} & = & \left[B_{\mu}^{2}\pm e_{3}B_{\mu}^{3}\right];\label{eq:46}\\
Z^{0} & = & \frac{e_{0}g_{m}B_{\mu}^{0}+e_{1}g_{m}^{'}B_{\mu}^{1}}{\sqrt{g_{m}^{2}+g_{m}^{'^{2}}}}.\label{eq:47}\end{eqnarray}

\textbf{Case - III }For magneto - electroweak unification i.e.$SU(2)_{e}\times U(1)_{m}$,
we assume $g_{m}^{'}=g_{e}=0$ and the mass contribution to Lagrangian
density is thus given by

\begin{eqnarray}
L_{mass} & = & \left(v\frac{g_{e}^{'}}{2}\right)^{2}W_{\mu}^{+}W_{-}^{\mu}+\frac{v^{2}}{8}\left(\begin{array}{cc}
B_{\mu}^{0} & A_{\mu}^{1}\end{array}\right)\left(\begin{array}{cc}
g_{m}^{2} & g_{m}g_{e}^{'}\\
g_{m}g_{e}^{'} & g_{e}^{'2}\end{array}\right)\left(\begin{array}{c}
B_{\mu}^{0}\\
A_{\mu}^{1}\end{array}\right)\label{eq:48}\end{eqnarray}
which leads to the charged $W$ boson masses as,

\begin{eqnarray}
m_{W} & = & \frac{v}{2}g_{e}^{'};\label{eq:49}\\
W_{\mu}^{\pm} & = & A_{\mu}^{2}\pm e_{3}A_{\mu}^{3};\label{eq:50}\\
Z^{0} & = & \frac{e_{1}g_{e}^{'}A_{\mu}^{1}+e_{0}g_{m}B_{\mu}^{0}}{\sqrt{g_{m}^{2}+g_{e}^{'2}}}.\label{eq:51}\end{eqnarray}

\textbf{Case - IV }For electro-magnetoweak unification i.e. $SU(2)_{m}\times U(1)_{e}$,
we assume $g_{e}^{'}=g_{m}=0$ and the mass contribution to Lagrangian
density is given by

\begin{eqnarray}
L_{mass} & = & \left(v\frac{g_{m}^{'}}{2}\right)^{2}W_{\mu}^{+}W_{-}^{\mu}+\frac{v^{2}}{8}\left(\begin{array}{cc}
A_{\mu}^{0} & B_{\mu}\end{array}\right)\left(\begin{array}{cc}
g_{e}^{2} & g_{e}g_{m}^{'}\\
g_{e}g_{m}^{'} & g_{m}^{'2}\end{array}\right)\left(\begin{array}{c}
A_{\mu}^{0}\\
B_{\mu}^{1}\end{array}\right)\label{eq:52}\end{eqnarray}
which leads to the charged $W$ boson masses as,

\begin{eqnarray}
m_{W} & = & \frac{v}{2}g_{m}^{'};\label{eq:53}\\
W_{\mu}^{\pm} & = & B_{\mu}^{2}\pm e_{3}B_{\mu}^{3};\label{eq:54}\\
Z^{0} & = & \frac{e_{1}g_{m}^{'}B_{\mu}^{1}+e_{0}g_{e}A_{\mu}^{0}}{\sqrt{g_{e}^{2}+g_{m}^{'2}}}.\label{eq:55}\end{eqnarray}
From the fore going analysis we find that the discovery of the spontaneous
symmetry breaking (SSB) and the Higgs mechanism in the non-Abelian
gauge theories describe a great break through towards the unification
of electromagnetic and weak interactions for different coupling parameters
of quaternion gauge theories. We may also conclude that the zeroth
model of quaternion leads the usual theory of electroweak interactions
of standard model. On the other hand the imaginary units quaternion
enlarges the gauge group leading to various gauge bosons which play
an important role in extra dimensions of string theory. As such, the
real understanding of the mechanism of the spontaneous breakdown and
the Higgs mechanism is still extremely challenging problem to be solved
in field theories.

\textbf{Acknowledgement: }One of us OPSN is thankful to Professor
H. Dehnen, Universität Konstanz, Fachbereich Physik, Postfach-M 677,
D-78457 Konstanz, Germany for his kind hospitality at Universität
Konstanz. He is also grateful to German Academic Exchange Service
(Deutscher Akademischer Austausch Dienst), Bonn for their financial
support under DAAD re-invitation programme.

\end{document}